# SALI: AN EFFICIENT INDICATOR OF CHAOS WITH APPLICATION TO 2 AND 3 DEGREES OF FREEDOM HAMILTONIAN SYSTEMS

Chris G. Antonopoulos[*], Athanasios E. Manos[*] and Charalampos D. Skokos[†*]

[*]Department of Mathematics
and
Center for Research and Applications of Nonlinear Systems (CRANS),
University of Patras,
26500 Patras, Greece
e-mail: antonop@math.upatras.gr, web page: http://www.math.upatras.gr/~antonop

[†]Research Center for Astronomy
Academy of Athens,
Anagnostopoulou 14, 10673 Athens, Greece

**Keywords:** Hamiltonian systems, Chaotic motion, Ordered motion, Lyapunov Exponents.

**Abstract.** *The **Smaller Alignment Index (SALI)** is a very useful and efficient indicator that can distinguish rapidly and with certainty between **ordered and chaotic motion** in Hamiltonian systems. This is based on the different behavior of the SALI in the two cases, as it fluctuates around non-zero values for ordered orbits and converges exponentially to zero for chaotic orbits. In this paper we present a detailed numerical study, comparing the advantages of the SALI to those of the **Maximal Lyapunov Exponent (MLE)**, which is the traditional indicator of chaos detection in dynamical systems. Exploiting these advantages, we demonstrate how the SALI can be used to identify even tiny regions of order and chaos in Hamiltonian systems of 2 and 3 degrees of freedom.*

## 1  INTRODUCTION

The distinction between ordered and chaotic motion in dynamical systems is fundamental in many areas of applied sciences. This distinction is particularly difficult in systems with many degrees of freedom (dof), basically because it is not feasible to visualize their phase space. So, we need fast and accurate tools to give us information about the chaotic or ordered character of orbits, especially for conservative systems.

Many methods have been developed over the years trying to give an answer to this problem. The inspection of the successive intersections of an orbit with a Poincaré surface of section (PSS)[1] has been used mainly for 2 dimensional (2d) maps and 2 dof Hamiltonian systems. One of the most common methods of chaos detection is the computation of the maximal Lyapunov Characteristic Number (LCN)[2,3], which can be applied for systems with many degrees of freedom. Another very efficient method is the "Frequency Map Analysis" developed by Laskar[4,5]. In recent years new methods have been introduced such as the study of spectra of "Short Time Lyapunov Characteristic Numbers"[6,7] or "Stretching Numbers"[8,9] and the "Spectral Distance" of such spectra[10], as well as the study of spectra of helicity and twist angles[11-13]. In addition, Froeschlé C., et al.[14] introduced the Fast Lyapunov Indicator (FLI), while Vozikis et al.[15] proposed a method based on the frequency analysis of "Stretching Numbers".

Recently, a new, fast and easy to compute indicator of the chaotic or ordered nature of orbits, has been introduced[16] and was applied successfully in different dynamical systems[16-23], the so-called "Smaller Alignment Index (SALI)". In the present paper, we first recall the definition of the Smaller Alignment Index and then we show its effectiveness in distinguishing between ordered and chaotic motion, by applying it to Hamiltonian systems of 2 and 3 dof.

## 2  DEFINITION OF THE SMALLER ALIGNMENT INDEX (SALI)

Let us consider the $n$-dimensional phase space of a conservative dynamical system, which could be a symplectic map or a Hamiltonian flow. We consider also an orbit in that space with initial condition $P(0)=(x_1(0),x_2(0),\ldots,x_n(0))$ and a deviation vector $\xi(0)=(dx_1(0),dx_2(0),\ldots,dx_n(0))$ from the initial point $P(0)$. In order to compute the SALI for a given orbit one has to follow the time evolution of the orbit with initial condition $P(0)$ itself and also of two deviation vectors $\xi_1(t)$, $\xi_2(t)$ which initially point in two different directions. The evolution of these deviation vectors is given by the variational equations for a flow and by the tangent map for a discrete-time system. At every time step the two deviation vectors $\xi_1(t)$ and $\xi_2(t)$ are normalized and the SALI is then computed as

$$SALI(t) = \min\left\{\left\|\frac{\xi_1(t)}{\|\xi_1(t)\|} + \frac{\xi_2(t)}{\|\xi_2(t)\|}\right\|, \left\|\frac{\xi_1(t)}{\|\xi_1(t)\|} - \frac{\xi_2(t)}{\|\xi_2(t)\|}\right\|\right\} \qquad (1)$$

The properties of the time evolution of the SALI clearly distinguish between ordered and chaotic motion as follows: In the case of Hamiltonian flows or $n-$dimensional symplectic maps with $n \geq 2$, the SALI fluctuates around a non-zero value for ordered orbits, while it tends to zero for chaotic orbits[16-20]. In the case of 2d maps the SALI tends to zero both for ordered and chaotic orbits, following however completely different time rates[16], which again allows us to distinguish between the two cases.

In 2 and 3 dof Hamiltonian systems the distinction between ordered and chaotic motion is easy because the ordered motion occurs on a 2d or 4d torus respectively on which any initial deviation vector becomes almost tangent after a short transient period. In general, two different initial deviation vectors become tangent to different directions on the torus, producing different sequences of vectors, so that SALI does not tend to zero but fluctuates around positive values[19]. On the other hand, for chaotic orbits, any two initially different deviation vectors tend to coincide in the direction defined by the nearby unstable manifold and hence either coincides with each other, or become opposite. This means that the SALI tends to zero when the orbit is chaotic and to a non-zero value when the orbit is ordered. Thus, the completely different behavior of the SALI helps us to distinguish between ordered and chaotic motion in Hamiltonian systems with 2 and 3 dof and in general in dynamical systems of higher dimensionality.

## 3   APPLICATION OF THE SMALLER ALIGNMENT INDEX
### 3.1  The Hénon-Heiles 2 dof Hamiltonian system

In order to illustrate the effectiveness of the SALI method in the determination of the dynamical character of orbits (e.g. chaotic or ordered), we consider the 2 dof Hénon-Heiles Hamiltonian:

$$H(x, y, p_x, p_y) = \frac{1}{2}(p_x^2 + p_y^2) + \frac{1}{2}(x^2 + y^2) + x^2 y - \frac{1}{3}y^3 \equiv E \qquad (2)$$

where $E = 1/8$ is the constant total energy of the system, $x$, $y$ are the generalized coordinates and $p_x$, $p_y$ are the generalized conjugate momenta[25].

For the system (2) we consider three characteristic individual orbits and compute for each one the evolution of the SALI. The first orbit with initial condition $(x, y, p_x, p_y) = (0, 0.1, 0.490578, 0)$ we call it R, the second orbit with initial condition $(x, y, p_x, p_y) = (0, -0.0135, 0.499816, 0)$ we call it S and finally the orbit with initial condition $(0, -0.015, 0.499773, 0)$ we call it C. We also use $\xi_1(0) = (1, 0, 0, 0)$ and $\xi_2(0) = (0, 0, 1, 0)$ as initial deviation vectors.

As we can see from figure 1(a) the SALI of the orbit R (solid line) fluctuates around $SALI \approx 1.25$ remaining almost constant after $10^5$ time steps indicating the ordered character of the orbit, while the SALI of orbit C (dash dot line) falls abruptly reaching the limit of the accuracy of the computer precision ($10^{-16}$) after about 1000 time steps. This implies that the two deviation vectors are represented by the same or opposite sign numbers in the computer and thus they have been aligned with the direction of the most unstable nearby manifold. On the other hand, the SALI of orbit S (dot line) shows an intermediate behaviour. This orbit is called "sticky" as it remains at the borders of an island of stability for a long time but eventually enters the chaotic sea. Thus, the SALI shows the chaotic character of the orbit after about 5000 time steps when the "stickiness" phenomenon stops to play the dominant role in the dynamics. After this time, the SALI shows again the same behaviour as for the chaotic orbit C.

In figure 1(b) we see the evolution of the MLE of the same three orbits as a function of time (in logarithmic scale). The computation of MLE does not help us to decide with certainty the nature of the orbit as fast as the SALI. This becomes evident by comparing the evolution of the MLE and the SALI in figure 1. We can be absolutely certain that orbit S is chaotic at $t \simeq 10^4$ where $SALI \approx 10^{-16}$ or even at $t \simeq 5000$ where $SALI \approx 10^{-6}$. On the other hand, the corresponding MLE initially decrease and only after $t \simeq 5000$ start to show a deviation from this decrease, although, in order to get a clear evidence of convergence to a non-zero value, further computations are needed. Thus, using the SALI method we are able to determine the nature of orbits with great certainty and in general earlier than the computation of the corresponding MLEs.

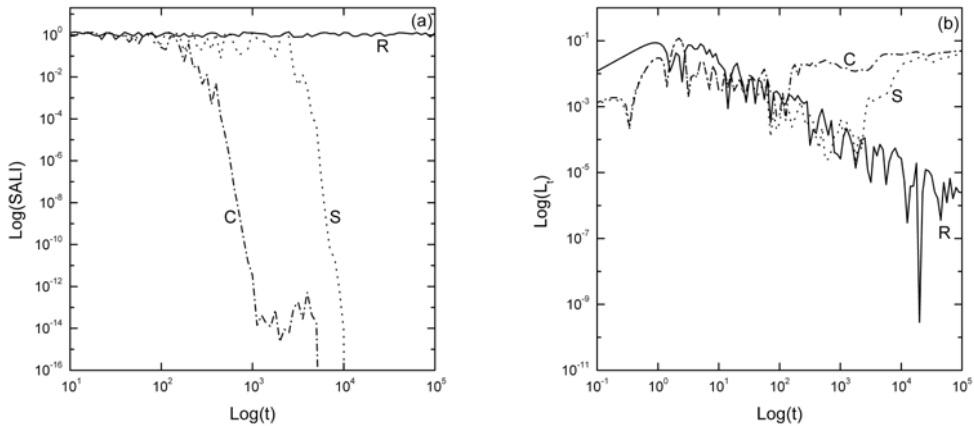

Figure 1. Evolution of (a) the SALI and (b) the MLE with respect to time $t$ in log-log scale for the orbits named R, S and C (see text).

In order to present the effectiveness of the SALI in detecting even tiny regions of chaos and order, we compute the SALI for a very large grid of equally distributed initial conditions on the PSS $(y, p_y)$ of Hamiltonian (2), assigning a different color to each initial condition according to the computed value of the SALI (we refer to this procedure as the full PSS scanning). In this way, we obtain a clear picture of the dynamics of the system on the PSS as we see in figure 2.

In particular, in figure 2(a) we compute the SALI of every individual initial condition on the grid for $t = 2000$ and assign an appropriate color according to the value of the index: If SALI$<10^{-12}$ the initial condition is colored black, if SALI $\in [10^{-12}, 10^{-8})$ the initial condition is colored dark gray, if SALI $\in [10^{-8}, 10^{-4})$ the initial condition is colored gray and finally, if SALI $\in [10^{-4}, \sqrt{2}]$ the initial condition is colored light gray. In producing figure 2(b), we performed a slightly different procedure to speed up the whole process without having significant errors: we assign the final value of the SALI at $t = 2000$ and the corresponding color not only to the initial condition but also to every subsequent intersection of the orbit with the PSS taking also into account the symmetry of the Hamiltonian (2) with respect to $p_y$ (we refer to this procedure as the fast PSS scanning). In this way, orbits with initial conditions on the grid to which a color has already been assigned, as they were intersection points with the PSS of previously computed orbits, are not computed again and so the whole process is speeded up significantly. In figure 2, we see that the differences of the two panels are almost not visible. We note that following the above procedure we are able to distinguish very tiny islands of stability located inside the chaotic sea.

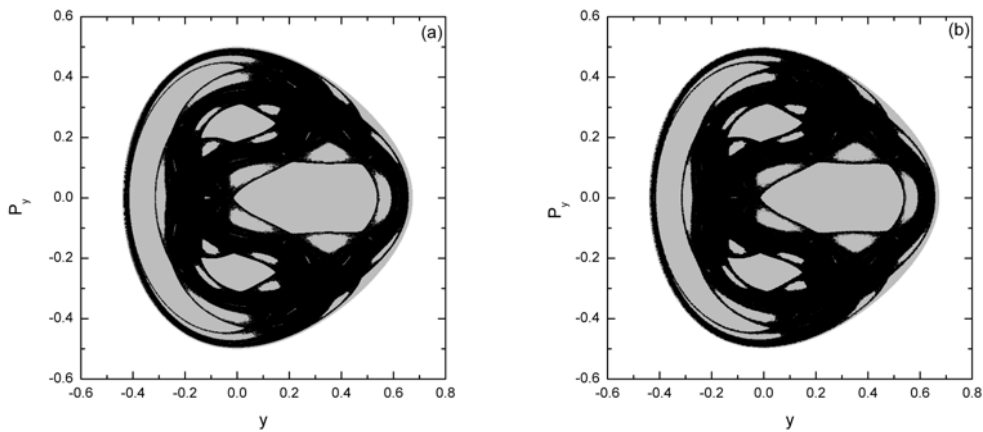

Figure 2. Regions of different values of the SALI on the PSS $x = 0$ of the Hénon-Heiles system (2) for a grid of $500 \times 500$ equally spaced initial conditions. In (a) the SALI was computed for every initial condition (full PSS scanning) while in (b) the final value of the SALI is attributed not only to the initial condition but also to all successive intersection points of the orbits with the PSS.

By applying the above procedure and assuming that an orbit is chaotic beyond any doubt if $SALI < 10^{-8}$, we computed the percentage of the chaotic orbits in the whole PSS. Similar results can be found in the paper of Hénon and Heils[25] where the authors used a different numerical technique in finding the percentage of chaotic orbits. The results of the full and fast PSS scanning are presented in figure 3. In figure 3(a), we see that both algorithms indicate that for energy $E = 1/8$ the PSS of figure 2 consists of about 55% of chaotic orbits as for $t > 2000$ both curves stabilize close to each other and further computations do not practically change the results. In figure 3(b) we plot the corresponding evolution of the CPU time on a Pentium 4 2GHz PC (in sec) needed for assigning a value of the SALI to every grid point on the PSS of system (2) as a function of the total time $t$ needed to integrate all the individual orbits. Practically, this is the CPU time needed for obtaining the data of figure 2(b). The CPU time increases for small integration times $t$ but after $t \simeq 2000$ exhibits a plateau as time $t$ is long enough for assigning a SALI value to many grid points, that we do not take as new initial conditions, since the computed orbits have many intersections with the PSS. As $t$ increases more, the CPU time slightly grows due to the increase of the integration time.

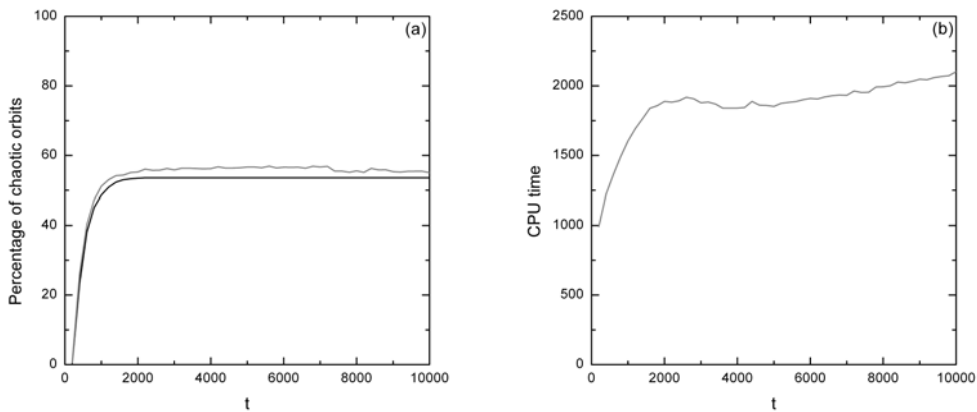

Figure 3. (a): The percentage of the chaotic orbits (having $SALI<10^{-8}$) with initial conditions on the PSS of Hamiltonian (2) for $E = 1/8$ as a function of the integration time $t$. (b): The evolution of the CPU time (in sec) needed for the fast scanning of the PSS. In panel (a) black line corresponds to the full scanning of the PSS while grey lines of both panels to the fast PSS scanning.

So, we realize that although the fast PSS scanning is not absolutely accurate (the two curves of figure 3(a) do not coincide) it is preferred for the complete study of big regions of the PSS as the CPU time required is significantly less than the CPU time needed for the full PSS scanning. For example, the full PSS scanning algorithm needs about 68300 sec of CPU time to integrate all the initial conditions on the grid for $t = 10000$, while the fast PSS scanning algorithm needs only about 2100 sec!

The SALI method can be used also to discriminate between order and chaos, not only on the PSS of a system, but also in any parameter space. As an example, in figure 4, we plot the SALI values on the plane $(y, E)$ for the Hamiltonian (2). In particular, we take orbits with initial conditions $x = 0$, $y \in [-0.3, 0.4]$, $p_y = 0$ and $E \in [0.08, 0.18]$ and calculate their SALI values for $t = 2000$. This is a similar plot to the one obtained by Fouchard et. al[26] using the FLI method for chaos detection. From figure (4) we see that as energy $E$ increases from 0.08 to 0.18 initial conditions with $y \in [-0.3, 0.4]$ that were initially ordered became chaotic, indicating that a lot of KAM tori (islands of stability) have been destroyed and in their position chaotic regions have been appeared. We stop our computations for $E = 0.18$ as for $E > 0.18$ escapes are possible.

**3.2 The relation between the SALI and the Lyapunov exponents in a 3 dof Hamiltonian system**

In order to illustrate the relation between the SALI and the 2 MLEs $L_1$ and $L_2$ respectively, we consider the 3 dof Hamiltonian system[27]

$$H(q_1, q_2, q_3, p_1, p_2, p_3) = \sum_{i=1}^{3} \frac{\omega_i}{2}(q_i^2 + p_i^2) + q_1^2 q_2 + q_1^2 q_3 \equiv E \qquad (3)$$

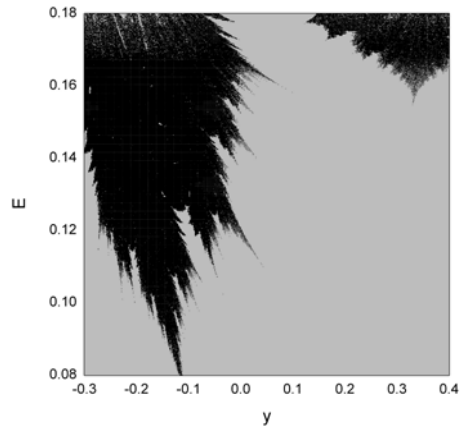

Figure 4: The SALI values for a grid of $500 \times 500$ equally spaced initial conditions in the plane $(y, E)$ for the Hénon-Heiles system (2) each integrated up to $t = 2000$. If SALI<$10^{-12}$ the initial condition is colored black, if SALI $\in [10^{-12}, 10^{-8})$ the initial condition is colored dark grey, if SALI $\in [10^{-8}, 10^{-4})$ the initial condition is colored grey and if SALI $\in [10^{-4}, \sqrt{2}]$ the initial condition is colored light grey.

where $q_1, q_2$ and $q_3$ are the generalized coordinates, $p_1, p_2$ and $p_3$ are the generalized conjugate momenta and $E = 0.09$ is the total constant energy of the system. Following [27], we use in our calculations $\omega_1 = 1, \omega_2 = 1.4142$ and $\omega_3 = 1.7321$ for compatibility reasons. The characteristic property of system (3) according to [27] is that the system possesses two kinds of chaotic regions in its phase space. It has a large stochastic region where the two MLEs are of the same magnitude, called the "big sea" and stochastic regions, which are apparently separated from the "big sea" called "small seas". In these "small seas" the two MLEs are not of the same magnitude and they differ significantly.

According to the recently published paper by Skokos et. al[20], the evolution of the SALI is related to the two largest Lyapunov exponents $L_1$ and $L_2$ of a 3 dof Hamiltonian system by

$$SALI(t) \propto e^{-(L_1 - L_2)t}. \quad (4)$$

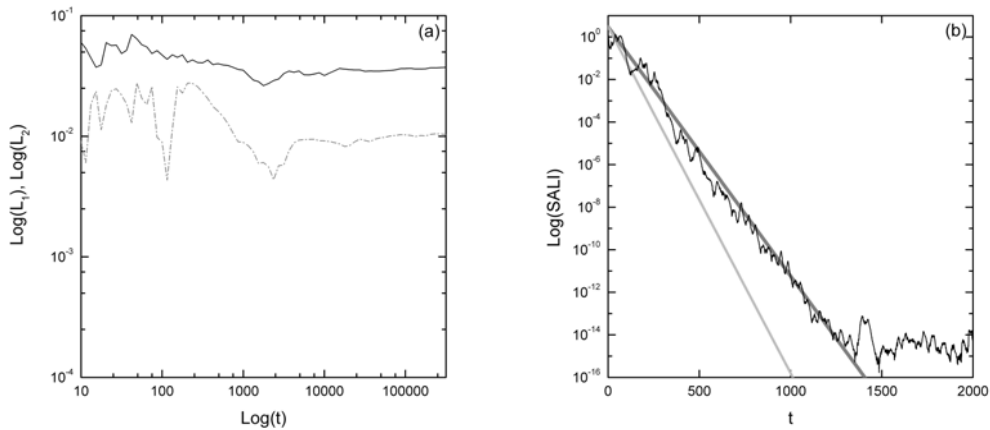

Figure 5. (a): The evolution of the two largest Lyapunov exponents $L_1(t)$ and $L_2(t)$ for the chaotic orbit with initial condition $(q_1, q_2, q_3, p_1, p_2, p_3)$=(0,0,0,0.14142,0.11892,0.29428) of the 3 dof Hamiltonian (3). (b): Time evolution of the SALI for the same orbit (black solid line) and the functions $ce^{-(L_1 - L_2)t}$ (grey line) and $ce^{-L_1 t}$ (light grey line) for $L_1 = 0.03747$, $L_2 = 0.0105$ and $c = 3$. Note that the $t$-axis is linear.

In order to support the validity of equation (4), we consider the orbit of system (3) with initial condition $(q_1, q_2, q_3, p_1, p_2, p_3)$=(0,0,0,0.14142,0.11892,0.29428) originated at the "big stochastic sea". In figure 5(a) we plot the time evolution of the two largest Lyapunov exponents $L_1(t)$ and $L_2(t)$ of the orbit which are computed by the method proposed by Benettin et. al.[28]. We note that at final time $t \approx 3 \cdot 10^5$ both exponents are of the same order of magnitude as $L_1 \approx 0.03747$ and $L_2 \approx 0.0105$. In figure 5(b), we plot the SALI of the orbit as a function of time $t$ (in linear scale) as well as the approximated functions $ce^{-(L_1-L_2)t}$ and $ce^{-L_1 t}$ for $L_1 = 0.03747$, $L_2 = 0.0105$ and $c = 3$. We see that using the SALI method we determine the chaotic nature of the orbit much faster than the computation of the Lyapunov exponents as for $t \simeq 600$ we have $\text{SALI} \simeq 10^{-8}$ while $L_1(t)$ and $L_2(t)$ still do not show clear evidence of convergence. We also note that the second Lyapunov exponent $L_2$ plays a crucial role in describing the behaviour of the SALI, as $e^{-(L_1-L_2)t}$ approximates much better the values of the SALI than $e^{-L_1 t}$ (see figure 5(b)).

## 4  CONCLUSIONS

In this paper we have illustrated the advantages of the SALI method in distinguishing between order and chaos in 2 and 3 dof Hamiltonian systems. Our results can be summarized as follows:

- The SALI proves to be an ideal indicator of chaoticity independent of the dimensions of the system. It tends to zero for chaotic orbits while it fluctuates around non-zero positive values for ordered ones, distinguishing clearly between the two cases.
- The SALI values characterize an orbit of being chaotic or ordered. Exploiting this feature of the index, we have plotted detailed phase space portraits for a 2 dof Hamiltonian system where the chaotic and ordered regions are clearly distinguished. We were thus able to trace very small islands of ordered motion, whose detection by traditional methods would be much more difficult and time consuming.
- We have also presented an example of a chaotic orbit in a 3 dof Hamiltonian system, whose SALI tends to zero following an exponential rate which is related to the difference of the two largest Lyapunov exponents $L_1$ and $L_2$. In this particular case, although $L_2$ is smaller than $L_1$, both exponents have the same order of magnitude and so, $L_2$ cannot be neglected from the description of the time evolution of the SALI.

## 5  ACKNOWLEDGMENTS

Ch. Antonopoulos was partially supported by the "Karatheodory" graduate student fellowship No 2464 of the University of Patras and by the "Heraclitus" research program of the Greek Ministry of Development. Ch. Skokos was partially supported by the Research Committee of the Academy of Athens.